\documentstyle[aps,prl]{revtex}

\input psfig.sty

\begin{document}
\twocolumn

\title{Vortices in a Bose-Einstein Condensate}
\author{M.~R. Matthews, B.~P. Anderson,\cite{qpdNIST} P.~C. Haljan, D.~S. Hall,\cite{amherst}  C.~E.
  Wieman, and E.~A. Cornell\cite{qpdNIST}}
\address{JILA, National Institute of Standards and Technology and Department of Physics, \\
University of Colorado, Boulder, Colorado 80309-0440}
\date{August 9, 1999}
\maketitle

\begin{abstract}
We have created vortices in two-component Bose-Einstein
condensates. The vortex state was created through a coherent
process involving the spatial and temporal control of
interconversion between the two components. Using an interference
technique, we map the phase of the vortex state to confirm that it
possesses angular momentum. We can create vortices in either of
the two components and have observed differences in the dynamics
and stability.

\end{abstract}

\pacs{03.75.Fi,67.90.+z,67.57.Fg,42.50.Md}

\par
The concept of a vortex is at the center of our understanding of
superfluidity. A vortex is a topological feature of a superfluid
-- in a closed path around the vortex, the phase undergoes a
$2\pi$ winding and the superfluid flow is quantized. Following the
experimental realization of a dilute atomic Bose-Einstein
condensate (BEC) \cite{reviews}, much theoretical effort has been
directed towards the formation and behavior of vortices in atomic
BEC \cite{2statetheory,theory,bigelow1999}. This paper presents
the experimental realization and imaging of a vortex in BEC. We
use the method proposed by Williams and Holland
\cite{Williams1999b} to create vortices in a two-component BEC. An
interference technique is used to obtain phase images of the
vortex state and confirm the $2\pi$ phase winding required by the
quantization condition.  We have also carried out preliminary
studies of the stability of the vortices.
\par
Vortices are created in superfluid helium by cooling a rotating bucket of
helium through the superfluid transition, and a vortex forms for each unit of
angular momentum. This does not work for BEC because it is formed in a harmonic
magnetic trap.  When the condensate first forms it occupies a tiny
cross-sectional area at the center of the trap and is too small to support a
vortex. Eventually, the condensate grows to sufficient size that it can support
vortices, but the time scale for vortices to be generated in the vortex-free
condensate due to coupling with the rotating environment is unknown, and may
well be longer than the lifetime of the condensate.  This is the potential
difficulty with using an optical ``stirring beam'' or magnetic field distortion
to rotate the cloud during condensation, as has been frequently proposed.
Another proposal has been to use optical beams with appropriate topologies to
``imprint'' a phase on an existing condensate. This technique must drive the
local density to zero at some point and then rely on uncertain dissipative
processes for the condensate to relax into a vortex state.
\par
We have avoided these uncertainties by creating vortices using a
coherent process that directly forms the desired vortex wave
function via transitions between two internal spin states of $
^{87}$Rb. The two spin states, henceforth \protect{$\left | 1
\right>$} and \protect{$\left | 2 \right>$}, are separated by the
ground-state hyperfine splitting and can be simultaneously
confined in identical and fully overlapping magnetic trap
potentials. A two-photon microwave field induces transitions
between the states. As we have seen in previous experiments, this
coupled two-component condensate is exempt from the topological
rules governing single-component superfluids \cite{Matthews1999a}
- rules that make it difficult to implant a vortex within an
existing condensate in a controlled manner.  In the coupled
system, we can directly create a \protect{$\left | 2 \right>$} (or
\protect{$\left | 1 \right>$}) state wave function having a wide
variety of shapes \cite{Williams1999b} out of a \protect{$\left |
1 \right>$} (or \protect{$\left | 2 \right>$}) ground-state wave
function by controlling the spatial and temporal dependence of the
microwave-induced conversion of \protect{$\left | 1 \right>$} into
\protect{$\left | 2 \right>$}.
\par
\begin{figure}[p]
\begin{center}
\psfig{figure=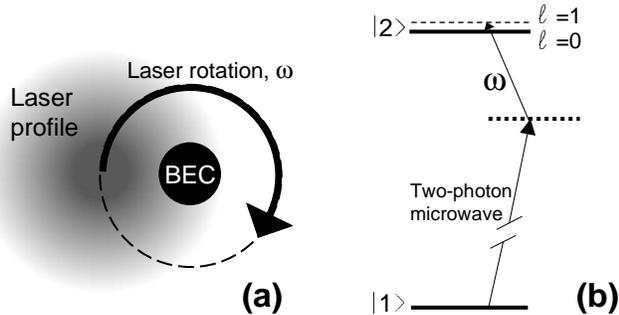,width=1\linewidth,clip=}
\end{center}

\caption {(a) A basic schematic of the technique used to create a
vortex. An off-resonant laser provides a rotating gradient in the
AC Stark shift across the condensate as a microwave drive of
detuning $\delta$ is applied. (b) A level diagram showing the
microwave transition to very near the \protect{$\left | 2
\right>$} state, and the modulation due to the laser rotation
frequency that couples only to the angular momentum $l=1$ state
when $\omega=\delta$.  In the figure, the energy splitting ($< 1$
Hz) between the $l=1$ and $l=0$ states is
exaggerated.}\label{beam}

\end{figure}

We control the conversion by shifting the transition frequency
using the AC Stark effect. A spatially inhomogeneous and moveable
optical field (a focused laser beam) provides the desired spatial
and temporal control of the AC Stark shift. The vortex state is an
axially symmetric ring with a $2\pi$ phase winding around the
vortex core where the local density is zero. To create a wave
function with this spatial symmetry, the laser beam is rotated
around the initial condensate as in Fig.~\ref{beam}a. The desired
spatial phase dependence is obtained by detuning the microwave
frequency from the transition, and rotating the laser beam at the
appropriate frequency $\omega$ to make the coupling resonant. For
large microwave detunings $\delta$, the necessary rotation
frequency is simply $\delta$. For smaller detunings, the rotation
frequency must be the effective Rabi frequency of the microwave
transition \cite{coupling}. As shown in Fig.~\ref{beam}b for large
detunings, the energy resonance condition now means that atoms can
only change internal state through the coupling of the
time-varying perturbation, and are therefore obliged to obey any
selection rule that the spatial symmetry of that perturbation
might impose. The center of the condensate (at the axis of the
beam rotation) feels no time-varying change, while regions near
the circumference of the condensate feel a near-sinusoidal
variation, with a phase delay equal to the azimuthal angle
$\theta$ around the circumference of the cloud. Williams and
Holland show that this is precisely the geometry best suited to
couple the condensate into a vortex state. It should be emphasized
that it is not simply the mechanical forces of the optical field
that excite the vortex: a laser beam rotating clockwise can
produce clockwise or counterclockwise vortex circulation,
depending on the sign of the microwave detuning.

\par
In the absence of the microwave coupling field, the two states can
be thought of as two distinguishable, interpenetrating superfluids
that interact with each other and with themselves via a mean-field
repulsion proportional to the local densities. The interaction
coefficients differ slightly \cite{Matthews1998a,Burke1998a}, so
the \protect{$\left | 1 \right>$} fluid has slight positive
buoyancy with respect to the \protect{$\left | 2 \right>$} fluid
\cite{Hall1998a}. When the \protect{$\left | 1 \right>$} fluid has
a net angular momentum, it forms an equatorial ring around the
central \protect{$\left | 2 \right>$} fluid. The \protect{$\left |
1 \right>$} fluid partially penetrates the constant-phase
\protect{$\left | 2 \right>$} fluid, which creates a central
potential barrier. Conversely, a \protect{$\left | 2 \right>$}
vortex forms a ring that tends to contract down into the
\protect{$\left | 1 \right>$} fluid. We use the overlap of the
\protect{$\left | 1 \right>$} and \protect{$\left | 2\right>$}
fluids to image the phase profile of the vortex state via the
interconversion interference technique that we introduced in
\cite{Hall1998b}. In the presence of a near-resonant microwave
field (and no perturbing optical field), the two states
interconvert at a rate sensitive to the local difference in the
quantum phases of the two states. Thus the application of a
resonant $\pi/2$ microwave pulse transforms the original two-fluid
density distribution into a distribution that reflects the local
phase difference, a ``phase interferogram.'' Looking at the
condensate both before and after the interconversion pulse
provides images of both the amplitude and phase of the vortex
ring.
\par
The basic experimental setup for forming condensates and driving
them between different spin states is the same as in
\cite{Matthews1998a}. Using laser cooling and trapping, followed
by trapping in a TOP magnetic trap and evaporative cooling, we
produce a condensate of typically $\sim8\times10^5$ atoms in the
\protect{$\left | 1 \right>$} state ($F=1$, $m_F=-1$). We then
adiabaticaly convert the trap to a spherically symmetric potential
by reducing the quadrupole magnetic field gradient
\cite{ensherthesis}. This leaves us with a condensate 54 microns
in diameter in a trap with oscillation frequencies of 7.8 $\pm$0.1
Hz in the radial and axial directions for both spin states. In
this trap, a \protect{$\left | 1 \right>$} state condensate has a
lifetime of 75 s and the \protect{$\left | 2 \right>$} state
($F=2$, $m_F=+1$) about 1 s. Oscillating magnetic fields can then
be pulsed on to drive the microwave transition between the
\protect{$\left | 1 \right>$} and \protect{$\left | 2 \right>$}
states. The power and oscillation frequency of the fields are
adjusted to obtain the desired effective Rabi frequency for the
\protect{$\left | 1 \right>$} to \protect{$\left | 2 \right>$}
transition. To create a vortex in \protect{$\left | 2 \right>$} we
add a 10 nW, 780 nm laser beam that has a waist of 180 $\mu$m and
is detuned 0.8 GHz blue of resonant excitation of the
\protect{$\left | 2 \right>$} state. Using piezoelectric actuators
we rotate the beam in a $\sim$75 $ \mu$m radius circle around the
condensate at 100 Hz. The following procedure is used to obtain
the precise location of the laser beam that is required to create
a vortex. We first set the effective Rabi frequency to 100 Hz by
adjusting the detuning of the microwave frequency away from
resonance in the presence of the light. Typically the resonant
Rabi frequency is 35 Hz and the total detuning about 94 Hz from
the \protect{$\left | 1 \right>$} to \protect{$\left | 2 \right>$}
transition. By making fine adjustments of the detuning we then
optimize the amount of transfer (typically 50\%) to the
\protect{$\left | 2 \right>$} state. We then adjust the center of
rotation of the beam to obtain the most symmetric rings. After
$\sim70$ ms the vortex has been formed, and we turn off the laser
beam and the microwave drive.
\par
We can take multiple images of a vortex both during and after
formation using nondestructive state-selective phase contrast
imaging \cite{Matthews1999a,Andrews}. Rapid control of the
microwave power and frequency allows us to apply various pulse
sequences to explore many different options for the creation,
manipulation, and observation of a single vortex. For example, we
can put the initial condensate into either the \protect{$\left | 1
\right>$} or \protect{$\left | 2 \right>$} state and then make a
vortex in the \protect{$\left | 2 \right>$} or \protect{$\left | 1
\right>$} state respectively, and we can obtain phase
interferograms or quickly switch the internal state of the vortex
at any time after the rotating laser beam is off. We can also
watch the evolution of the vortex over time scales from
milliseconds to seconds.  All of these techniques are
nondestructive to both the density and phase.
\par
\begin{figure*}[p]
\begin{center}
\psfig{figure=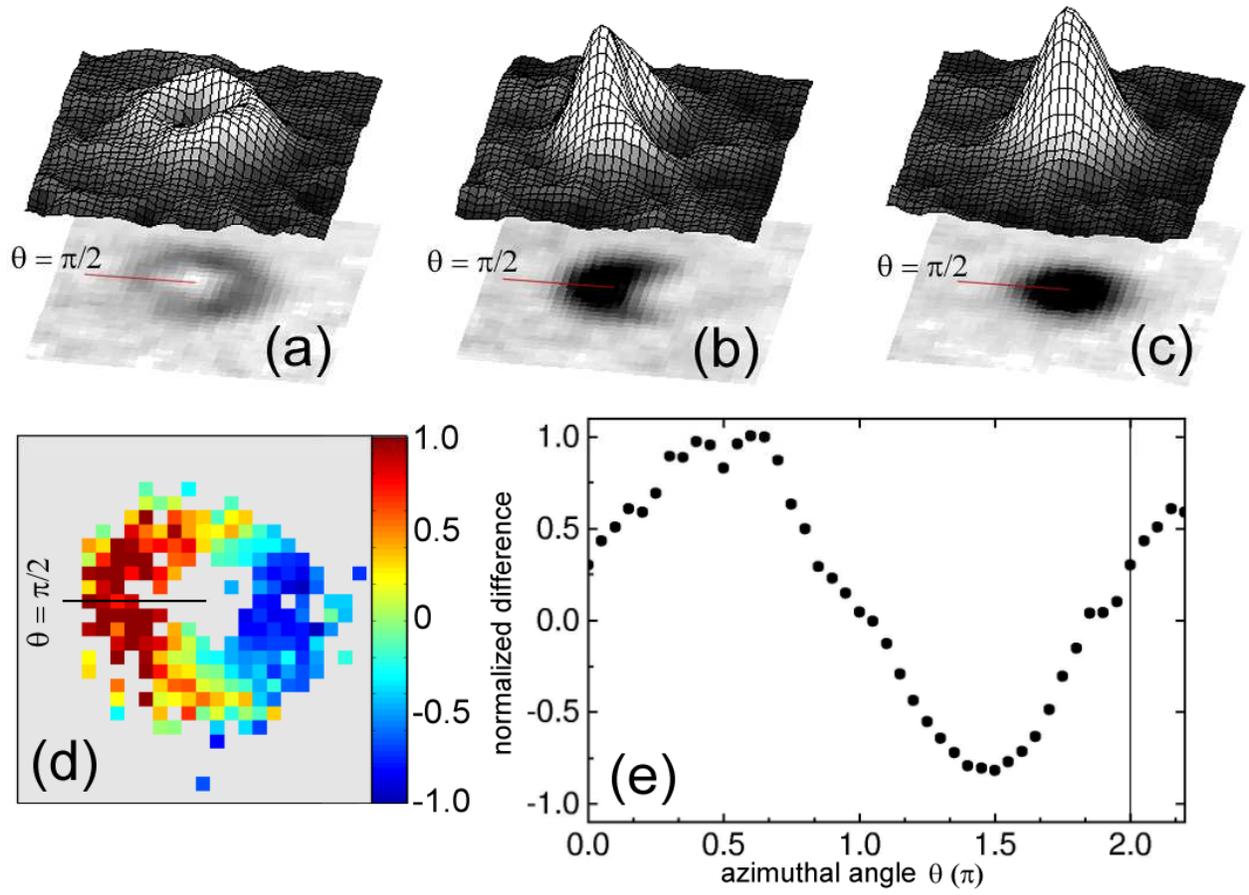,width=1\linewidth,clip=1}
\end{center}
\caption {Condensate images (100 $\mu$m on a side) in which the
imaging laser is tuned such that only \protect{$\left | 2
\right>$} is visible. (a) The intial image of the vortex, (b)
after $\pi/2$ of the interconversion pulse, and (c) after
completion of the $\pi$-pulse. The vortex is now invisible (in
\protect{$\left |1 \right>$}) and the interior fluid is imaged (in
\protect{$\left |2 \right>$}). (d) The normalized difference in
densities between the local average ring and interior densities
((a) and (c) respectively) and the phase interferogram density (b)
for each corresponding point in the images. This is approximately
the cosine of the local phase difference $\phi$ between the vortex
state and the interior state [14]. The values are shown only for
those regions where the densities of each state were high enough
to give adequate signal to noise for the phase reconstruction. (e)
The radial average at each angle $\theta$ around the ring is shown
in (d). (The data is repeated after the azimuthal angle $2\pi$ to
better show the continuity around the ring.)} \label{phase}
\end{figure*}

In Fig. \ref{phase} we show a detailed picture of the phase
profile of a \protect{$\left | 2 \right>$}vortex.  To obtain this
we first take a picture of the vortex (Fig. \ref{phase}a), then we
apply a resonant microwave $\pi$-pulse. Half-way through the 4 ms
long $\pi$-pulse, we take a second image (Fig. \ref{phase}b), and
at the completion of the pulse we take a third image (Fig.
\ref{phase}c) that shows the original density distribution of the
interior \protect{$\left | 1 \right>$} state. Normalizing by the
density distributions of the vortex and interior states we obtain
the phase image in Fig. \ref{phase}d \cite{fracdens}. The figure
dramatically shows the variation and continuity of the phase
around the ring (Fig. \ref{phase}e) that are required by the
quantization of angular momentum.

\onecolumn

\twocolumn

\par
\begin{figure}[p]
\begin{center}
\psfig{figure=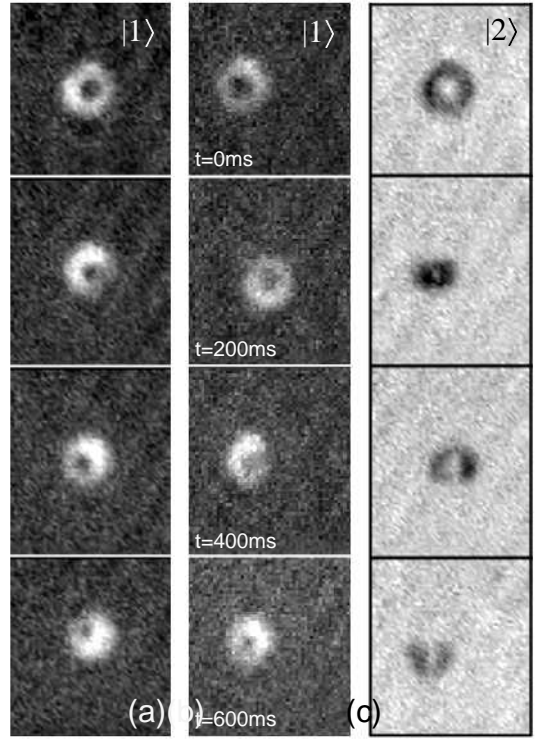,width=1\linewidth,clip=1}
\end{center}
\caption {(a,b) Two separate instances of the free evolution of a
\protect{$\left | 1\right>$} state vortex in the magnetic trap. It
is stable over a time long compared to the trap oscillation period
(128 ms).  (c) The free evolution of a \protect{$\left | 2
\right>$} vortex is much more dynamic. It is seen shrinking
quickly into the invisible \protect{$\left | 1\right>$} fluid and
rebounding into fragments. Each column is from a single run, where
time $t$ is referenced to the end of vortex creation ($t$ is the
same for each row).  The \protect{$\left | 1 \right>$} and
\protect{$\left | 2 \right>$} state images appear different due to
different signs of the probe detuning.} \label{time}
\end{figure}

In Fig. \ref{time} are pictures of the time evolution of vortices
to show their dynamics and stability.  As expected, the dynamics
of the \protect{$\left | 1 \right>$} state vortices (Fig.
\ref{time}a and Fig. \ref{time}b) are different from the
\protect{$\left | 2 \right>$} state vortices (Fig. \ref{time}c)
due to the different scattering lengths.  The \protect{$\left | 2
\right>$} vortex ring sinks in toward the trap center (Fig.
\ref{time}c), and then rebounds and apparently breaks up. This
pattern is repeatedly observed in measurements of the
\protect{$\left | 2 \right>$} vortices.  Conversely, the
equilibrium position of a \protect{$\left | 1 \right>$} state
fluid is obtained by ``floating'' outside the \protect{$\left | 2
\right>$} state fluid. The inner radius of the \protect{$\left | 1
\right>$} state ring shrinks slowly as the interior
\protect{$\left | 2 \right>$} fluid decays away with a $\sim 1$~s
lifetime.  A variety of additional behaviors has been seen for
\protect{$\left | 1 \right>$} vortices.  Initial asymmetry is very
sensitive to beam position and condensate slosh. For small
differences in the initial vortex state density distribution,
asymmetric density distributions sometimes develop and/or ``heal''
between 0.5 and 1 s.

\par
The nonrotating \protect{$\left | 2 \right>$} fluid within a
\protect{$\left | 1 \right>$} vortex is analogous to the defects
that pin vortices in superconductors.  This  \protect{$\left | 2
\right>$} ``defect'' can be removed quickly, by a properly tuned
laser pulse, or allowed to decay slowly (as in Figs. \ref{time}a
and \ref{time}b). Thus this system can be varied between two
relevant physical limits. In the limit of a large repulsive
central potential (produced by a large amount of  \protect{$\left
| 2 \right>$} fluid)  the system most closely resembles quantized
flow in a fixed, 3-D toroidal potential. The vortex core is pinned
and its size is determined by the central potential.  Figs.
\ref{time}a and \ref{time}b are near this limit for $t$= 0 and 200
ms. In this case the density distribution of the vortex state is
vulnerable to instabilities due to the fact that the relative
densities of \protect{$\left | 1 \right>$} and \protect{$\left | 2
\right>$} may evolve with relatively little energy cost so long as
the total density remains constant \cite{Hall1998a}. In the
opposite limit (a small amount of \protect{$\left | 2 \right>$}
fluid), the central potential is negligible and the size of the
vortex core (in equilibrium) is determined entirely by its own
centrifugal barrier. Figs. \ref{time}a and \ref{time}b at $t$ =
600 ms are evolving towards this limit.  In Fig. \ref{time}a, we
see that the interior \protect{$\left | 2 \right>$} fluid is no
longer pinning the vortex at 600 ms, but rather is being dragged
around by the precessing vortex. The \protect{$\left | 2 \right>$}
vortex provides an interesting mixture of the above limits.  The
\protect{$\left | 1 \right>$} fluid ``floats'' to the outside so
there is no pinning of the vortex core, but the tendency towards
density instabilities remains.
\par
Expanded studies of stability issues are underway.  We also expect
to be able to observe interesting transitional behavior between
these limiting cases. For example, it is a straightforward
extension of our method to create an $l=2$ vortex. In the presence
of a strong pinning potential, $l=2$ vortices should be stable,
but in the weak potential limit, $l=2$ vortices are predicted to
spontaneously bifurcate \cite{bigelow1999}. We gratefully
acknowledge useful conversations with J. Williams and M. Holland.
This work is supported by the ONR, NSF, and NIST.

\begin{thebibliography}{10}
\bibitem{coupling}
In the limit of large detuning, $\delta$ is equal to the effective Rabi
frequency.  In our implementation, $\omega$ and the effective Rabi frequency
are 100 Hz and $\delta$ is 94 Hz.
\bibitem{fracdens}
The normalized difference is defined as $(2d_{\left |1 \right> \left |2
  \right>}- d_{\left |2 \right>}-d_{\left |1 \right>})/(2\sqrt{d_{\left
  |1\right>} d_{\left |2 \right>}}$) where $d_{\left |i \right>}$ is the local
  density for the $i$th state and $d_{\left |1 \right> \left |2 \right>}$ for the
  interferogram (Fig. \ref{phase}b). This is not exactly the cosine of the phase difference
  $\phi$ between the vortex and interior state due to small effects, such as incomplete
  overlap between the states along the line of sight and uncertainties in the zero levels.
\end{thebibliography}
\end{document}